\documentclass[reqno]{amsart}
\usepackage{latexsym,amsmath,amsfonts,graphics,amssymb}

\begin{document}
\title{Connectivity in one-dimensional ad hoc networks with an access point}

\author[J. Li]
{Junshan Li}  

\address{Institute of Applied Mathematics, East China Jiao Tong University, Nanchang, China.}
\email{junshanli@ecjtu.org.cn}

\begin{abstract}
In this paper, we study the connectivity in one-dimensional ad hoc
wireless networks with an fixed access point. In recent years,
various closed expressions for the probability of connectivity on
one-dimensional networks (interval graphs) have been derived by many
researchers. We will provide some numerical validation for them by
means of extensive simulations.

\bigskip

\textbf{Keywords:} interval graph; ad hoc network; connectivity;
component.
\end{abstract}

\bigskip

\maketitle

\numberwithin{equation}{section}
\newtheorem{theorem}{Theorem}[section]
\newtheorem{lemma}[theorem]{Lemma}
\newtheorem{proposition}[theorem]{Proposition}
\newtheorem{definition}[theorem]{Definition}
\newtheorem{corollary}[theorem]{Corollary}
\newtheorem*{remark}{Remark}

\section{Introduction}

Recently, ad hoc networks have attracted extensive research interest
within computer communication and engineering communities.
Connectivity of the underlying network is the foundation for its
functions, as is indicated in \cite{9,8}. A number of mathematically
rigorous results on the asymptotic critical transmission radius and
asymptotic critical neighbors for the connectivity of network in
one-dimensional areas have been obtained, see e.g.
\cite{13,4,7,5,6,12,11,3}. The random version of one-dimensional
network is also called the random interval graph, which has been
studied in depth in \cite{1}. More recently, a concept of  access
points are introduced in \cite{10}. Given $L,r>0$, let
$X_1,\cdots,X_n$ be $n$ independent uniformly distributed random
variables on the interval $[0,L]$. Denote by $G(n,L)$ the graph with
vertex set $\{X_1,\cdots,X_n\}$ and with an edge $X_iX_j$
($i\not=j$), if $|X_i-X_j|\le r$. Some access points $\{z_j\}$ can
exist in $G(n,L)$. An interesting concept of accessible connectivity
has also been introduced, which is different from the ordinary
connectivity with possible fixed nodes. Multiple components issues
are considered in \cite{12} ($cf.$ Theorem 2.2 below).

In this paper we will perform numerical study on the connectivity of
one-dimensional ad hoc networks with a fixed point. Our work can be
viewed as an independent test of the theoretical results obtained in
prior work.

\section{Numerical study}

We need some further definitions here. We denote the graph by
$G_x(n,L)$ if there exists a fixed node at the point $x\in[0,L]$. In
particular, when $x=0$, the graph is $G_0(n,L)$. We use $Q_m(\cdot)$
to denote the probability that the above graph model is composed of
exactly $m$ components. The following results are known.

\begin{theorem} (\cite{7,6}) We have
\begin{equation}
Q_1(G(n,L))=\sum_{i=0}^{k_1}(-1)^i{n-1 \choose
i}(1-\frac{ir}{L})^n,\label{1}
\end{equation}
and
\begin{equation}
Q_1(G_0(n,L))=\sum_{i=0}^{k_2}(-1)^i{n \choose
i}(1-\frac{ir}{L})^n,\label{2}
\end{equation}
where $k_1=n-1\wedge \lfloor L/r\rfloor$ and $k_2=n\wedge \lfloor
L/r\rfloor$, respectively.
\end{theorem}

More generally, the following statements are true.

\begin{theorem} (\cite{12}) Let $m$ be a natural number. We have
\begin{equation}
Q_m(G(n,L))=\sum_{i=m-1}^{k_1}(-1)^{i-m+1}{i \choose m-1}{n-1\choose
i}(1-\frac{ir}{L})^n,\label{3}
\end{equation}
and
\begin{equation}
Q_m(G_0(n,L))=\sum_{i=m-1}^{k_2}(-1)^{i-m+1}{i \choose m-1}{n\choose
i}(1-\frac{ir}{L})^n,\label{4}
\end{equation}
where $k_1=n-1\wedge \lfloor L/r\rfloor$ and $k_2=n\wedge \lfloor
L/r\rfloor$, respectively.
\end{theorem}

To test the results in Theorem 2.2, we have performed extensive
simulations on graph $G(n,L)$ with a fixed node at $0$. For
different values of $n$ and $m$, we plot the probability $Q_m$ as a
function of $L/r$, see Fig.1 --Fig. 6. A future research problem
would be to derive exact formula on two dimensional ad hoc networks.
This issue is very challenging.

\begin{figure}[htb]
\begin{center}
\scalebox{0.5}{\includegraphics{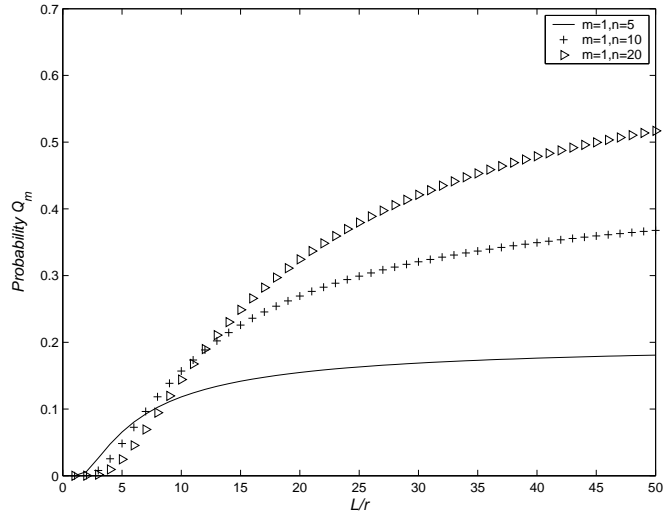}}\caption{The probability
$Q_m(G_0(n,L))$ as a function of $L/r$ for $m=1$ and  different
values of $n$: $n=5$ (solid curves), $n=10$ (pluses) and $n=20$
(triangles).}
\end{center}
\end{figure}

\begin{figure}[htb]
\begin{center}
\scalebox{0.5}{\includegraphics{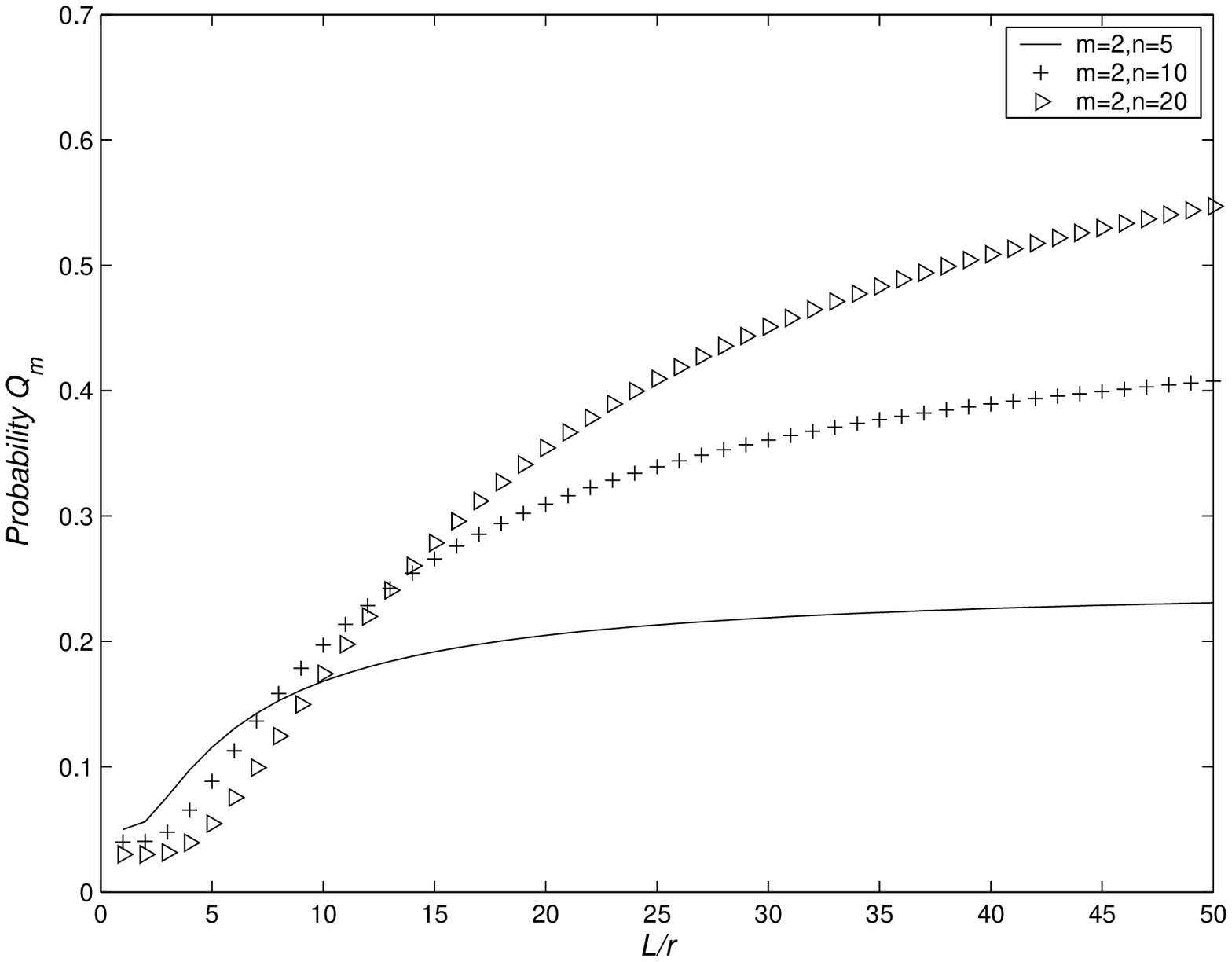}}\caption{The probability
$Q_m(G_0(n,L))$ as a function of $L/r$ for $m=2$ and  different
values of $n$: $n=5$ (solid curves), $n=10$ (pluses) and $n=20$
(triangles).}
\end{center}
\end{figure}

\begin{figure}[htb]
\begin{center}
\scalebox{0.5}{\includegraphics{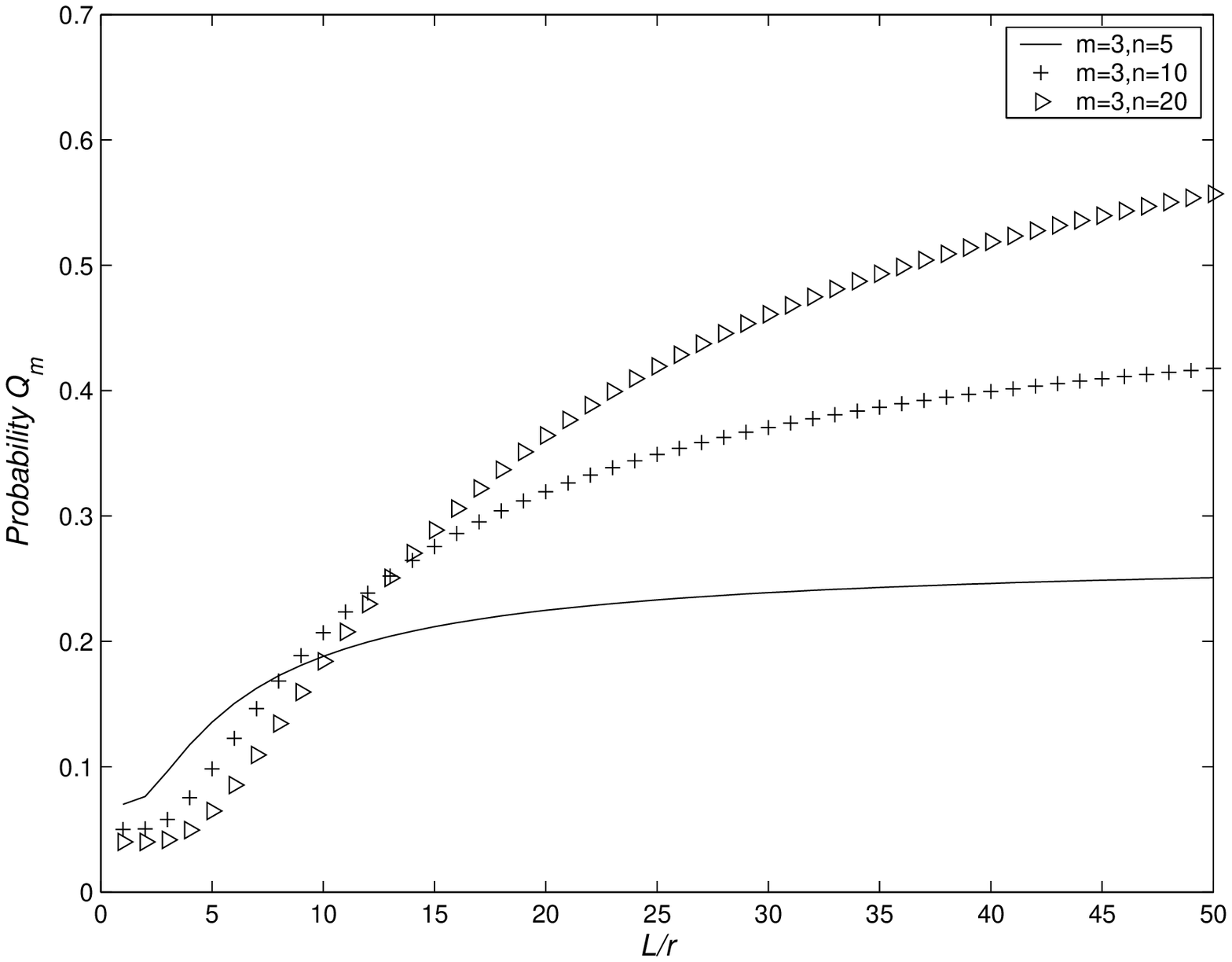}}\caption{The probability
$Q_m(G_0(n,L))$ as a function of $L/r$ for $m=3$ and  different
values of $n$: $n=5$ (solid curves), $n=10$ (pluses) and $n=20$
(triangles).}
\end{center}
\end{figure}

\begin{figure}[htb]
\begin{center}
\scalebox{0.5}{\includegraphics{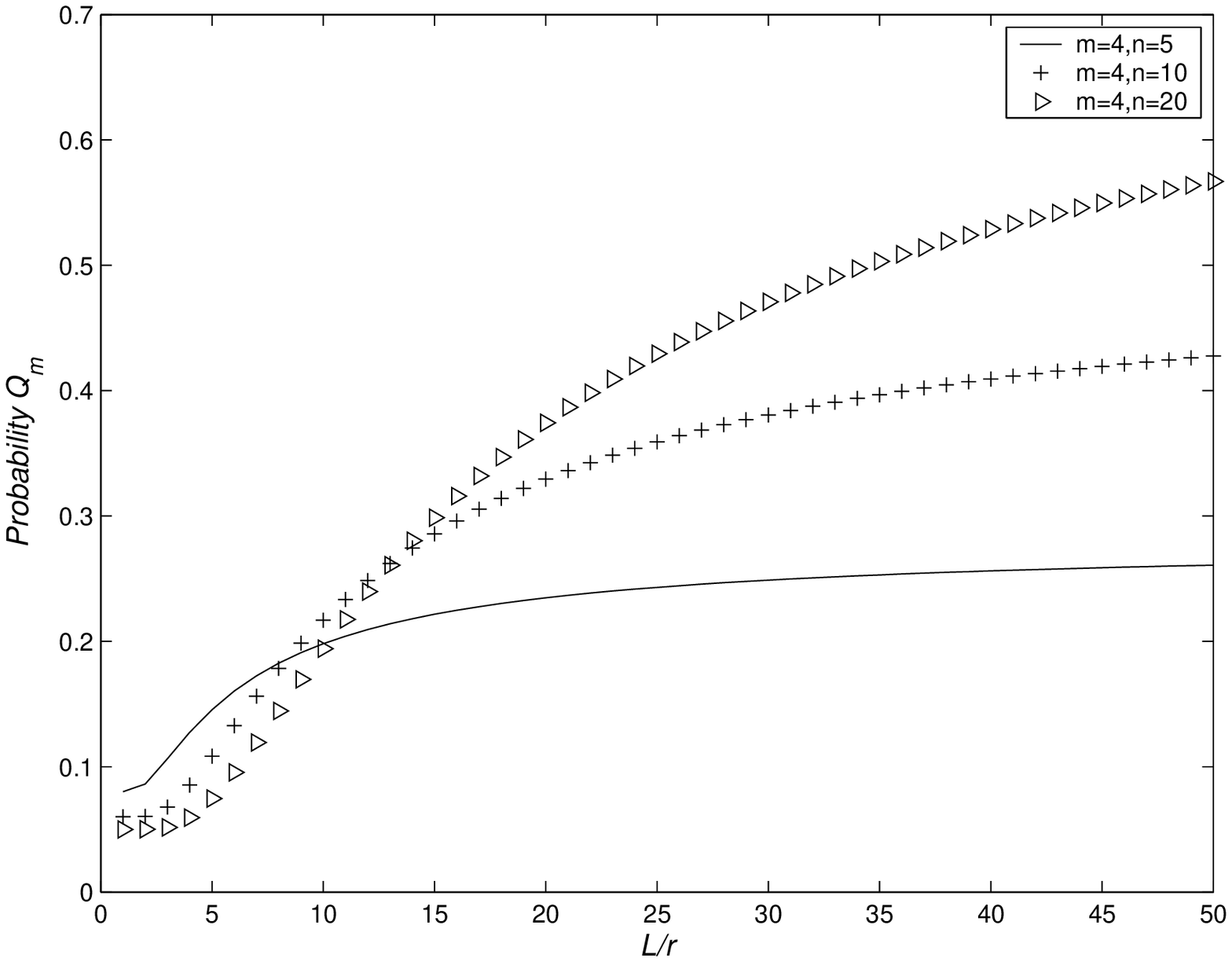}}\caption{The probability
$Q_m(G_0(n,L))$ as a function of $L/r$ for $m=4$ and  different
values of $n$: $n=5$ (solid curves), $n=10$ (pluses) and $n=20$
(triangles).}
\end{center}
\end{figure}

\begin{figure}[htb]
\begin{center}
\scalebox{0.5}{\includegraphics{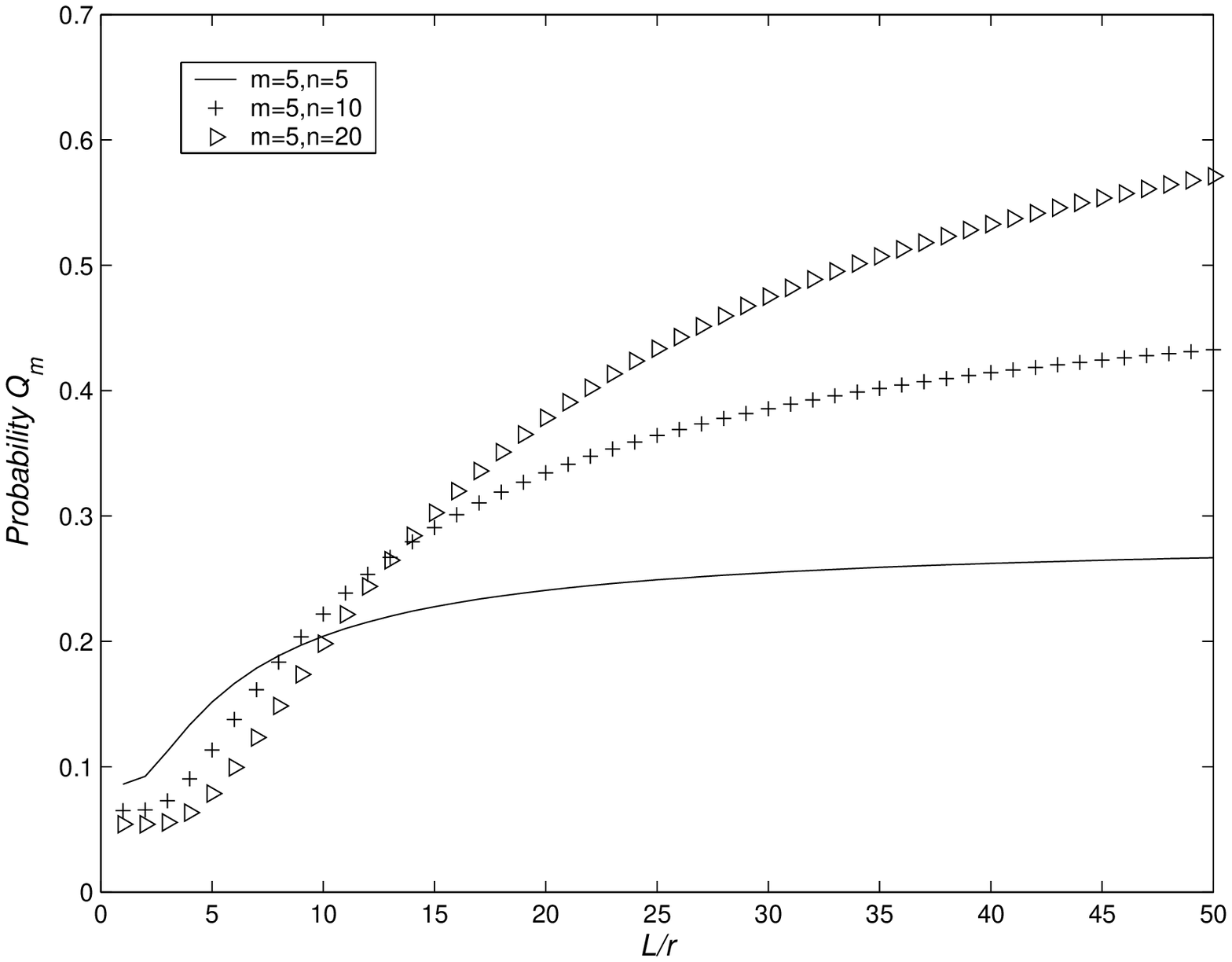}}\caption{The probability
$Q_m(G_0(n,L))$ as a function of $L/r$ for $m=5$ and  different
values of $n$: $n=5$ (solid curves), $n=10$ (pluses) and $n=20$
(triangles).}
\end{center}
\end{figure}

\begin{figure}[htb]
\begin{center}
\scalebox{0.5}{\includegraphics{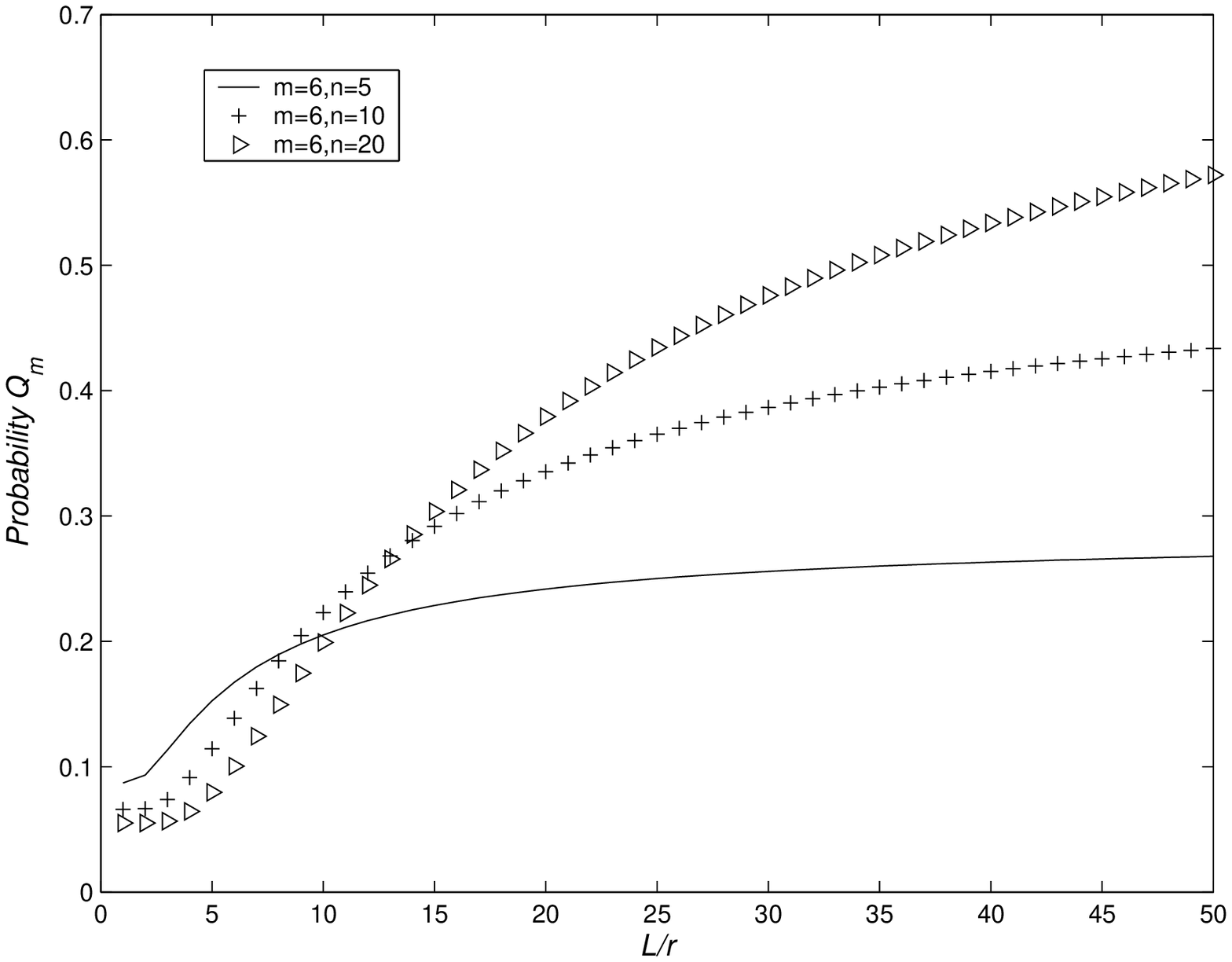}}\caption{The probability
$Q_m(G_0(n,L))$ as a function of $L/r$ for $m=6$ and  different
values of $n$: $n=5$ (solid curves), $n=10$ (pluses) and $n=20$
(triangles).}
\end{center}
\end{figure}

\end{document}